\documentclass[letterpaper, 10 pt, journal, twoside]{IEEEtran}
\IEEEoverridecommandlockouts                              

\usepackage{url}
\usepackage{cite}
\usepackage{amsmath,amssymb,amsfonts}
\usepackage{graphicx}
\usepackage{textcomp}
\usepackage[dvipsnames]{xcolor}
\usepackage{soul}
\usepackage{multirow}
\usepackage[mathscr]{euscript}
\usepackage[normalem]{ulem}
\usepackage[linesnumbered,ruled,vlined]{algorithm2e}
\SetKwInput{KwInput}{Input}                
\SetKwInput{KwOutput}{Output}

\def\BibTeX{{\rm B\kern-.05em{\sc i\kern-.025em b}\kern-.08em
    T\kern-.1667em\lower.7ex\hbox{E}\kern-.125emX}}

\makeatletter
\renewcommand*\env@matrix[1][*\c@MaxMatrixCols c]{%
  \hskip -\arraycolsep
  \let\@ifnextchar\new@ifnextchar
  \array{#1}}
\makeatother

\newcommand{\vo}[1]{\boldsymbol{#1}}
\newcommand{\mo}[1]{\boldsymbol{#1}}
\newcommand{\A}{\vo{A}}
\newcommand{\B}{\vo{B}}

\newcommand{\ddd}{\vo{d}}
\newcommand{\x}{\vo{x}}
\newcommand{\y}{\vo{y}}
\newcommand{\z}{\vo{z}}
\newcommand{\w}{\vo{w}}
\newcommand{\n}{\vo{n}}

\newcommand{\Lg}{\vo{L}} 

\newcommand{\nx}{N_x}
\newcommand{\ny}{N_y}
\newcommand{\nz}{N_z}

\newcommand{\nd}{N_d}
\newcommand{\ns}{N_{\Sens}}
\newcommand{\ks}{k_{\Sens}}

\newcommand{\precs}{p}
\newcommand{\scln}{\sigma}

\newcommand{\Sens}{\mathcal{S}}
\newcommand{\Htwo}{\mathcal{H}_2}
\newcommand{\Hinf}{\mathcal{H}_{\infty}}
\newcommand{\norm}[2]{\left\lVert#1\right\rVert_{#2}}
\newcommand{\fnorm}[1]{\norm{#1}{F}} 
\newcommand{\fprod}[2]{\langle #1, #2 \rangle _F} 

\newcommand{\Bd}{\vo{B}_d}

\newcommand{\Bw}{\vo{B}_w}

\newcommand{\Cy}{\vo{C}_y}
\newcommand{\Cz}{\vo{C}_z}
\newcommand{\Dd}{\vo{D}_d}

\newcommand{\Dw}{\vo{D}_w}

\newcommand{\Af}{\vo{A}_F}
\newcommand{\Bf}{\vo{B}_F}
\newcommand{\Cf}{\vo{C}_F}
\newcommand{\xf}{\vo{x}_F}

\newcommand{\Ae}{\vo{A}_E}
\newcommand{\Be}{\vo{B}_E}
\newcommand{\Ce}{\vo{C}_E}
\newcommand{\xe}{\vo{x}_E}

\newcommand{\zerr}{\vo{\varepsilon}}
\newcommand{\xerr}{\x_E}

\newcommand{\GO}{\vo{\mathcal{G}}_{O}(s)}
\newcommand{\GF}{\vo{\mathcal{G}}_{F}(s)}
\newtheorem{theorem}{Theorem}

\newtheorem{example}{Example} 
\newtheorem{problem}{Problem} 
\newtheorem{remark}{Remark} 
\newtheorem{definition}{Definition} 

\newcommand{\betab}{\vo{\beta}}

\newcommand{\xdot}{\dot{\vo{x}}}

\newcommand{\set}[1]{\mathcal{#1}}

\newcommand{\I}[1]{\vo{I}_{#1}}

\newcommand{\X}{\vo{X}}

\newcommand{\W}{\vo{W}}
\newcommand{\Y}{\vo{Y}}
\newcommand{\R}{\vo{R}}
\newcommand{\M}{\vo{M}}
\newcommand{\N}{\vo{N}}
\newcommand{\Q}{\vo{Q}}
\newcommand{\U}{\vo{U}}

\newcommand{\sym}[1]{\textbf{sym}\left(#1\right)}
\newcommand{\ith}{$i^{\text{th}}$ }

\renewcommand{\P}{\mo{P}} 
\newcommand{\Real}{\mathbb R}

\renewcommand{\vec}[1]{\textbf{vec}\left(#1\right)}

\newcommand{\inner}[1]{\left\langle \vo{e}\phi_i\right\rangle}

\newcommand{\eqnlabel}[1]{\label{eqn:#1}}

\newcommand{\eqn}[1]{(\ref{eqn:#1})}

\newcommand{\fig}[1]{Fig. (\ref{fig:#1})}

\DeclareMathAlphabet{\mathbfsf}{\encodingdefault}{\sfdefault}{bx}{n}

\newcommand{\Z}{\vo{Z}}

\newcommand{\C}{\vo{C}}
\newcommand{\D}{\vo{D}}

\renewcommand{\H}{\vo{H}}

\newcommand{\domain}[1]{\set{D}}

\newcommand{\diag}{\textbf{diag}}

\newcommand{\trace}[1]{\text{tr}\left( #1 \right)}

\title{Sensor Selection and Optimal Precision in $\Htwo/\Hinf$ Estimation Framework: Theory and Algorithms}

\author{Vedang M. Deshpande$^{1}$ and Raktim Bhattacharya$^{2}$
\thanks{This work was supported by the National Science Foundation (grant
number: 1762825).}
\thanks{$^{1}$Vedang M. Deshpande is a Ph.D. student in Aerospace Engineering, Texas A\&M University, College Station, TX 77843, USA. {\tt\small vedang.deshpande@tamu.edu}}%
\thanks{$^{2}$Raktim Bhattacharya is Associate Professor in Aerospace Engineering,
Electrical \& Computer Engineering, Texas A\&M University, College Station, TX 77843, USA. {\tt\small raktim@tamu.edu}}}

\begin{document}
\maketitle
\thispagestyle{empty} 
\begin{abstract}
   We consider the problem of sensor selection for designing observer and filter for continuous linear time invariant systems such that the sensor precisions are minimized, and the estimation errors are bounded by the prescribed $\Htwo/\Hinf$ performance criteria. The proposed integrated framework formulates the precision minimization as a convex optimization problem subject to linear matrix inequalities, and it is solved using an algorithm based on the alternating direction method of multipliers (ADMM). We also present a greedy approach for sensor selection and demonstrate the performance of the proposed algorithms using numerical simulations.
\end{abstract}

\section{INTRODUCTION}
Estimation of dynamical systems is an old yet rich problem in control and systems literature, and still an active area of research. The conventional problem of designing an estimator (observer or filter) involves determining the estimator parameters such as observer gain or filter matrices to achieve certain performance index for a given system with a \textit{pre-specified} set of sensors of \textit{known} precisions \cite{maybeck1982stochastic, kailath2000linear, el2000advances}. As a general rule, using all available sensors yields the best performance of an estimator.  However, doing this may not be always feasible due to various reasons such as economic budget limitations or weight and size restrictions on physical prototypes. Therefore, the problem of selecting a subset from available sensors has received significant attention especially in the last couple of decades
\cite{lopez_sparse_2014, yang_deterministic_2015, nugroho_simultaneous_2018, joshi_sensor_2009, chepuri_sparsity-promoting_2015, summers_submodularity_2016, manohar_optimal_2020, bopardikar_randomized_2021, shamaiah_greedy_2010, tzoumas_sensor_2016, zhang_sensor_2017, dhingra_admm_2014, zare_proximal_2020, jovanovic_controller_2016, munz_sensor_2014, hibbard_sensor_2020, li_integrating_2008, saraf_h2_2017, goyal_integrating_2020, deshpande_sparseH2Hinf_LCSS2021, deshpande_sparseRobHinf_ACC2021, das_sparseKalman_ICSSA2017, das_EnUKF_IFAC2020, das_mrKalman_2020, deshpande_sparseMultiAgent_ACC2021}
and has been applied to engineering systems such as power grids \cite{summers_submodularity_2016}, battery systems \cite{lin_robust_2020}, vibration control \cite{hiramoto_optimal_2000}, transportation systems \cite{contreras_observability_2016}, etc.


In this paper, we concern ourselves with the sensor selection problem at design-time, i.e. the set of sensors is chosen once and does not change over time. This is a combinatorial problem and becomes intractable even for systems of moderate sizes.
Various works by a number of researchers provide tractable alternatives for obtaining a (sub-)optimal solution to this problem using different approaches, for example, convex relaxations \cite{joshi_sensor_2009}, augmenting cost function with a sparsity promoting term \cite{dhingra_admm_2014, zare_proximal_2020, jovanovic_controller_2016, munz_sensor_2014}, greedy \cite{shamaiah_greedy_2010, tzoumas_sensor_2016, zhang_sensor_2017} and randomized \cite{bopardikar_randomized_2021} algorithms, and value iterations \cite{hibbard_sensor_2020}.

In \cite{joshi_sensor_2009, chepuri_sparsity-promoting_2015}, the cardinality constraint or $l_0$-norm of the sensor selection vector is replaced by its convex relaxation, i.e. $l_1$-norm, and heuristics are discussed to select a subset of sensors based on the solution of the relaxed problem. Sensor selection for non-linear
models with Cramér–Rao bound as a performance metric is discussed in \cite{chepuri_sparsity-promoting_2015}.

The problem of sensor selection to optimize a scalar measure of observability of dynamic networks is discussed in \cite{summers_submodularity_2016}. The authors consider different metrics of observability and prove the associated (sub/super-)modular properties to leverage the power of greedy algorithms to produce the optimal or sub-optimal solutions with guaranteed optimality bounds \cite{nemhauser_analysis_1978}.  A similar formulation \cite{manohar_optimal_2020} exploits balanced model reduction and greedy methods to optimize the observability for high-dimensional systems. A randomized algorithm to select sensors such that the observability Gramian is sufficiently non-singular by a user-specified margin, and probabilistic guarantees on the resulting solution are presented in \cite{bopardikar_randomized_2021}.

The optimal sensor placement problem for Kalman filtering is discussed in \cite{shamaiah_greedy_2010,tzoumas_sensor_2016, zhang_sensor_2017}. Works \cite{shamaiah_greedy_2010} and \cite{zhang_sensor_2017} aim to minimize certain measures of error covariance while satisfying a cardinality constraint on the sensor set, on the other hand, the formulation \cite{tzoumas_sensor_2016} seeks to minimize the number of sensors while guaranteeing a desired bound on the covariance. The sub/super-modularity of the log-determinant function is exploited in \cite{shamaiah_greedy_2010,tzoumas_sensor_2016} to guarantee the well-known $(1-1/e)$ optimality of  greedy algorithms \cite{nemhauser_analysis_1978}. The authors of \cite{zhang_sensor_2017} showed that the sensor selection problem they considered is NP-hard, but does not exhibit sub/super-modular structure for a general system.

In \cite{dhingra_admm_2014, zare_proximal_2020, jovanovic_controller_2016, munz_sensor_2014} and related papers, the objective is to minimize $\Htwo/\Hinf$ norm of the estimator error system. To promote sparsity in the sensor configuration, the objective function is augmented with a weighted penalty on the columns of the observer gain matrix. The optimization problems are solved efficiently using the customized algorithms based on alternating direction method of multipliers (ADMM) \cite{dhingra_admm_2014} and proximal gradient \cite{zare_proximal_2020}.

The aforediscussed sensor selection frameworks \cite{lopez_sparse_2014, yang_deterministic_2015, nugroho_simultaneous_2018, joshi_sensor_2009, chepuri_sparsity-promoting_2015, summers_submodularity_2016, manohar_optimal_2020, bopardikar_randomized_2021, shamaiah_greedy_2010, tzoumas_sensor_2016, zhang_sensor_2017, dhingra_admm_2014, zare_proximal_2020, jovanovic_controller_2016, munz_sensor_2014, hibbard_sensor_2020}
either completely disregard the sensor noises (and hence precisions) or assume that the sensor precisions are known and fixed.
Such choice of sensors with pre-specified precisions limit the performance of control and estimation algorithms. It is possible that a system could use sensors with unnecessarily high precisions for a desired level of performance, resulting in higher economic cost. In general, it is unclear,  which sensors should be improved or added in order to attain better performance by a specified margin. This problem becomes non-trivial for large-scale systems. In this paper, we treat sensor precisions as unknown design variables to be determined. Our goal is to design estimators that utilize \textit{minimum number} of sensors with \textit{minimum precisions} while guaranteeing certain performance criterion, thus, favoring economically cheaper physical systems.

An integrated framework for estimator/controller design in which sensor and actuator precisions are treated as variables was first introduced in \cite{li_integrating_2008}. They posed a convex optimization problem to incorporate linear constraints on precision variables arising due to economic budget limitations while guaranteeing steady-state covariance bounds in $\Htwo$ framework. Their work was extended for models with uncertainty in \cite{saraf_h2_2017}, and most recently, was applied to tensegrity systems in  \cite{goyal_integrating_2020}. In \cite{li_integrating_2008}, authors also presented an ad-hoc algorithm to achieve a sparse sensor configuration by iteratively eliminating a sensor with the least precision. In Section \ref{sec:sens_select}, we show that this algorithm produces solutions with arbitrarily large errors, making it unsuitable for high-dimensional systems.

Adopting the idea of variable sensor precisions from \cite{li_integrating_2008}, our recent work \cite{deshpande_sparseH2Hinf_LCSS2021} addressed the problem of observer design with given performance bound in both $\Htwo$ and $\Hinf$ frameworks while simultaneously minimizing the required sensor precisions and promoting sparseness in the sensor configuration. We also discussed an extension of this work for uncertain models in an $\Hinf$ framework in our most recent paper \cite{deshpande_sparseRobHinf_ACC2021}.  In \cite{deshpande_sparseH2Hinf_LCSS2021} and \cite{deshpande_sparseRobHinf_ACC2021}, we formulated a convex optimization problem as a semidefinite program (SDP) to minimize weighted $l_1$-norm of the precision vector subject to $\Htwo/\Hinf$  performance bounds written as linear matrix inequalities (LMIs). Standard software packages such as \texttt{CVX} \cite{grant_cvx_2020} were used to solve the SDP, and an iterative reweighting scheme \cite{candes_enhancing_2008} was used to promote sparsity in the sensor configuration. A Kalman filtering framework for the optimal precision problem with guaranteed error bounds has been presented in \cite{das_sparseKalman_ICSSA2017, das_EnUKF_IFAC2020} and extended for sensor scheduling applications in \cite{das_mrKalman_2020, deshpande_sparseMultiAgent_ACC2021}.

\subsubsection*{Contribution and novelty}
In this paper, we present an integrated theoretical framework to (i) design estimators (\textit{both observers and filters}) that satisfy the performance bounds specified in terms of $\Htwo$ or $\Hinf$ norm of the error system, (ii) select a subset of sensors that satisfies the given cardinality constraint and (iii) minimize the sensor precisions.

We extend the results from \cite{deshpande_sparseH2Hinf_LCSS2021} for \textit{filter design} in $\Htwo$ and $\Hinf$ estimation frameworks to simultaneously minimize the sensor precisions. The optimal precision problems are formulated as SDPs subject to LMIs. For such SDPs, the general-purpose SDP solvers do not scale well as the system size is increased \cite{dhingra_admm_2014, zare_proximal_2020}, therefore, we also present an ADMM algorithm that scales relatively well.

As discussed in the following section, sensor selection problem considered herein is different from \cite{deshpande_sparseH2Hinf_LCSS2021}, as in the present paper we require the selected subset of sensors to satisfy a hard cardinality constraint. We propose a greedy algorithm that iteratively solves the SDPs to arrive at a feasible subset of sensors that satisfies the cardinality constraint. Although the underlying function is shown not to exhibit submodular structure, empirical results show that the greedy sensor selection algorithm performs reasonably well in practice, and it is more accurate and reliable than the heuristics presented in \cite{li_integrating_2008} and the reweighting scheme in \cite{candes_enhancing_2008}.

\subsubsection*{Organization}
The paper is organized as follows. The sensor selection problem with optimal precision for estimator design is formulated in Section \ref{sec:prob}. The problem is split into two parts: (i) estimator design with optimal precision and (ii) sensor selection.
 The optimal precision problem for estimator design is discussed in Section \ref{sec:opt_prec} which also presents our main theoretical results and the ADMM algorithm to solve the optimization problem. The sensor selection aspect of the problem is discussed in Section \ref{sec:sens_select} along with the proposed greedy algorithm and its performance comparison with the heuristics from the literature. Section  \ref{sec:concl} presents the concluding remarks, and an appendix at the end provides solutions to the optimization sub-problems involved in the ADMM algorithm.

\subsubsection*{Notation} Unless specified otherwise, we adhere to the following notation throughout this paper.
The set of real numbers is denoted by $\Real$. Matrices (vectors) are denoted by bold uppercase (lowercase) letters. The transpose, trace and pseudo-inverse of a matrix $\X$ are respectively denoted by $\X^T$, $\trace{\X}$ and $\X^{\dagger}$.
For a square matrix $\X$, $\textbf{sym}\left(\X\right):=\X+\X^T$. The notation $\X>0$ ($\X<0$) is used for denoting  a symmetric positive (negative) definite matrix $\X$. Identity and zero matrices of suitable dimensions are denoted by $\I{}$ and $\vo{0}$ respectively.  Inequalities and exponents of vectors are to be interpreted elementwise. The diagonal matrix whose diagonal entries are a vector $\x$ is denoted by $\diag(\x)$, and the block-diagonal matrix with $N$ component matrices $\{\X_i\}_{i=1}^{N}$ is denoted by $\diag\left(\X_1,\X_2,\cdots,\X_N \right)$.

\section{Problem Formulation} \label{sec:prob}
Let us consider the following continuous linear time-invariant (LTI) system
\begin{align}
    \xdot(t) &= \A\x(t) + \Bd\ddd(t), \eqnlabel{sys_proc}
\end{align}
where, $\x\in\Real^{\nx}$ and $\ddd\in\Real^{\nd}$ are respectively the state vector and the process noise. We are interested in estimating a vector $\z\in\Real^{\nz}$ given by
\begin{align}
    \z(t) &= \Cz\x(t). \eqnlabel{sys_z}
\end{align}

We assume that we have been given a finite set of sensors $\Sens$ of cardinality $|\Sens|=:\ns$. The measurement equation for
every sensor $i\in\Sens$ is of the form
\begin{align}
    y_i(t) &= \C_{i}\x(t) +  \D_{i}\ddd(t) + \scln_i n_i(t), \eqnlabel{sys_meas_i}
\end{align}
where $y_i$ is a scalar measurement,  $\C_{i}$ and $\D_{i}$ are given row matrices of appropriate dimensions. The sensor measurement $y_i$ is corrupted by a noise signal $n_i$. The scalar $\scln_i >0$ is unknown and inversely related to the precision $\precs_i$ of the sensor, as discussed next.


The external disturbances $\ddd$ and $n_i$ are assumed to be stochastic signals ($\Htwo$ framework) or norm-bounded signals ($\Htwo/\Hinf$ frameworks).
When external disturbances are stochastic, they are assumed to be zero-mean stationary stochastic processes normalized to have unit signal variance.  Except $\scln_i$, all other normalizing weighting matrices are assumed to be known and absorbed in the system matrices. Precision of a sensor is defined to be the inverse of variance of the noise signal entering the sensor, therefore, $1/\scln_i^2=:\precs_i$ is the precision of the sensor $i$. In a similar manner, norm-bounded noise signals $n_i$ are assumed to be normalized to have unit energy ($\mathcal{L}_2$-norm), and sensor precision is defined to be the inverse of square of the signal energy, again yielding $\precs_i=1/\scln_i^2$ \cite{deshpande_sparseH2Hinf_LCSS2021}.

In the next section we will consider the problem of estimator design in $\Htwo/\Hinf$ frameworks to estimate $\z$.
As discussed in the introduction, our goal is to design estimators that use minimum number of sensors with minimum precisions, while satisfying an upper bound on $\Htwo/\Hinf$ norm of the associated error system. The number of sensors needed and their required precisions are, in general, competing interests, i.e. minimum precisions are achieved if all available sensors are used.
Therefore, we require that the selected subset of sensor must satisfy an upper bound on its cardinality.
The composite problem that we are considering in this work is stated as follows

\begin{problem} \label{prob:text}
  Given a set $\Sens$, weights $\rho_i>0, \forall i\in\Sens$, positive integer $\ks$, and $\gamma>0$, the objective is to (i) Select a subset $\set{Q}\subseteq\Sens$ such that $|\set{Q}|\leq\ks$, (ii) Minimize $\sum_{i\in\set{Q}}\rho_i p_i$, and  (iii) Design an estimator with the desired $\Htwo/\Hinf$ performance bound $\gamma>0$.
\end{problem}

 A brute force method to solve this problem would be to perform an exhaustive search over all subsets of $\Sens$ with cardinality less than or equal to $\ks$, and choose a subset that requires minimum sensor precisions. Since this is a combinatorial problem, the exhaustive search becomes intractable as the system dimension increases.
Therefore, we split the \textit{Problem \ref{prob:text}} in two parts. First, in Section \ref{sec:opt_prec},
we consider estimator design problem to minimize sensor precisions for a \textit{given subset} $\set{Q}\subseteq\Sens$. The second part to search over subsets of $\Sens$ using tractable algorithms until the cardinality constraint is satisfied, is discussed in  Section \ref{sec:sens_select}.


\section{Estimator Design with Optimal Sensor Precision} \label{sec:opt_prec}
In this section we focus on the problem of observer or filter design to minimize sensor precisions for a given set of sensors $\set{Q}$, such that $\set{Q}\subseteq\Sens$. Measurement equations \eqn{sys_meas_i} for all sensors $s_{l}\in\set{Q}$ can be written in a compact form as below
\begin{align}
    \y(t) &= \Cy\x(t) +  \Dd\ddd(t) + \diag(\vo{\scln})\n(t) , \eqnlabel{sys_meas}
\end{align}
where $\y:=[y_{s_1},y_{s_2}\cdots y_{s_{|\set{Q}|}}]^T\in\Real^{\ny}$ and $\ny:=|\set{Q}|$. Similarly,
the vectors $\vo{\scln},\n\in\Real^{\ny}$ are defined as $\vo{\scln}:=[\scln_{s_1} ,\scln_{s_2}\cdots \scln_{s_{|\set{Q}|}}]^T$, $\n:=[n_{s_1},n_{s_2}\cdots n_{s_{|\set{Q}|}}]^T$.
Finally, the measurement matrices are defined as $\Cy:=[\C_{s_1}^T \cdots \C_{s_{|\set{Q}|}}^T]^T\in\Real^{\ny\times\nx}$, $\Dd:=[\D_{s_1}^T \cdots \D_{s_{|\set{Q}|}}^T]^T\in\Real^{\ny\times\nd}$.

Let us also define the precision vector $\vo{\precs}:=1/\vo{\scln}^2$, and the given weights vector $0<\vo{\rho}:=[\rho_{s_1},\rho_{s_2}\cdots \rho_{s_{|\set{Q}|}}]^T\in\Real^{\ny}$. The objective for this part is to minimize the weighted $l_1$-norm of $\vo{\precs}$
$$\norm{\vo{\precs}}{1,\vo{\rho}}:=\sum_{i\in\set{Q}} \rho_i\precs_i.$$

We define an augmented vector of external disturbances $\w(t)$, and the associated matrices as follows
\begin{equation}
  \begin{aligned}
     \w(t) &:=\begin{bmatrix}\ddd^T(t) & \n^T(t)\end{bmatrix}^T, \\
       \Bw &:= \begin{bmatrix}\Bd & \vo{0}\end{bmatrix}, \ \Dw := \begin{bmatrix}\Dd & \diag(\vo{\scln})\end{bmatrix}, \eqnlabel{w_part}
  \end{aligned}
\end{equation}
which will be used in the text to follow.
We first consider the observer design problem.

\subsection{Observer Design}
Let us consider Luenberger observer of the form
\begin{equation}
\begin{aligned}
    \dot{\hat{\x}}(t) 
                      =& \left(\A+\Lg\Cy \right)\hat{\x}(t) - \Lg\y(t), \\
    \hat{\z}(t) =& \Cz\hat{\x}(t), \eqnlabel{obs}
\end{aligned}
\end{equation}
where $\hat{\x}\in\Real^{\nx}$ and $\hat{\z}\in\Real^{\nz}$  are estimates of the state $\x$ and the  output vector of interest $\z$, and the $\Lg \in \Real^{\nx \times \ny}$ is the unknown observer gain.
The estimation errors are defined as
\begin{align*}
    \xerr(t) := \x(t)-\hat{\x}(t), \text{ and }
    \zerr(t) := \z(t)-\hat{\z}(t). 
\end{align*}
The estimation error system follows from \eqn{sys_proc}, \eqn{sys_meas}, \eqn{w_part} and \eqn{obs}
\begin{equation}
\begin{aligned}
    \dot{\xerr}(t) &= \left(\A+\Lg\Cy\right)\xerr(t) + \left(\Bw+\Lg\Dw\right)\w(t),\\
    \zerr(t) &= \Cz\xerr(t).
\end{aligned} \eqnlabel{obs_err}
\end{equation}
Overall stability of the error system \eqn{obs_err} requires $\left(\A+\Lg\Cy \right)$  to be Hurwitz, or in other words, the pair $(\A,\Cy)$ must be detectable.
The transfer matrix of the system \eqn{obs_err} from $\w(t)$ to $\zerr(t)$  is given by
\begin{align*}
    \GO := & \Cz\big(s\I{}-\A-\Lg\Cy\big)^{-1} \big(\Bw +\Lg\Dw\big),
\end{align*}
where $s$ is the complex variable. The performance of an estimator is quantified in terms of $\Htwo$ or $\Hinf$ norm of the transfer matrix from external disturbances $\w(t)$ to the estimation error $\zerr(t)$. Therefore, the observer design problem with optimal precision for a given performance $\gamma >0$ is formally stated as follows
\begin{problem} \label{prob:obs}
  Minimize $\norm{\vo{\precs}}{1,\vo{\rho}}$ and determine a feasible $\Lg$, such that  $\norm{\GO}{\Htwo} < \gamma$ or $\norm{\GO}{\Hinf} <\gamma$.
\end{problem}

The solution to $\Htwo/\Hinf$ observer design \textit{Problem \ref{prob:obs}} has been discussed in detail in our previous work \cite{deshpande_sparseH2Hinf_LCSS2021}. For the sake of completeness and brevity, we adapt the results from \cite{deshpande_sparseH2Hinf_LCSS2021} for the system equations \eqn{sys_proc}, \eqn{sys_z}, \eqn{sys_meas}, and present them below without proofs.

\begin{theorem}[$\Hinf$ observer \cite{deshpande_sparseH2Hinf_LCSS2021}] \label{thm:hinf_obs}
The solution of $\Hinf$ observer design \textit{Problem \ref{prob:obs}} is determined by solving the following optimization problem, and the observer gain is given by $\Lg = \X^{-1}\Y$.
\begin{align}
\min\limits_{\vo{\precs}>0,\X>0,\Y}\quad \norm{\vo{\precs}}{1,\vo{\rho}}
\text{ such that } & \M(\vo{\precs},\X,\Y) < 0, \eqnlabel{hinf_thm_obs}
\end{align}
where
\begin{align}
& \M (\vo{\precs},\X,\Y) := \begin{bmatrix} \M_{11}  & \M_{12} & \Cz^T & \Y \\
               \ast & -\gamma\I{} & \vo{0}    & \vo{0} \\
                \ast &  \ast    & -\gamma\I{}     & \vo{0} \\
               \ast &  \ast    & \ast & -\gamma \ \diag(\vo{\precs}) \end{bmatrix}, \nonumber\\
& \M_{11}:=\sym{\X\A+\Y\Cy}, \M_{12}:=\X\Bd  +\Y\Dd.   \eqnlabel{hinf_obs_m11m12_def}
\end{align}

\end{theorem}

\begin{theorem}[$\Htwo$ observer \cite{deshpande_sparseH2Hinf_LCSS2021}] \label{thm:h2_obs}
The solution of $\Htwo$ observer design \textit{Problem \ref{prob:obs}} is determined by solving the following optimization problem, and the observer gain is given by $\Lg = \X^{-1}\Y$.
\begin{equation}\left.
\begin{aligned}
  & \min\limits_{\vo{\precs}>0,\X>0,\Y,\Q>0}\ \norm{\vo{\precs}}{1,\vo{\rho}}  \\
  & \text{ such that } \M(\vo{\precs},\X,\Y) < 0, \\
 & \begin{bmatrix} -\Q & \Cz \\ \Cz^T & -\X \end{bmatrix} < 0, \  \trace{\Q}<\gamma^2,
\end{aligned}\right\}\eqnlabel{h2_thm_obs}
\end{equation}

\begin{align*}
\text{where } \M (\vo{\precs},\X,\Y) := \begin{bmatrix} \M_{11} & \M_{12} & \Y \\
                    \ast & -\I{} & \vo{0} \\
                   \ast & \ast & -\diag(\vo{\precs}) \end{bmatrix},
\end{align*}
$\M_{11}:=\sym{\X\A+\Y\Cy}$ and $\M_{12}:=\X\Bd  +\Y\Dd$.
\end{theorem}

Next, we consider the filter design problem for a given set of sensors.

\subsection{Filter Design}
We assume the following form for the filter to estimate $\z(t)$
\begin{equation}
\begin{aligned}
    \dot{\x}_F(t) &= \Af\xf(t) + \Bf\y(t), \\
    \hat{\z}(t) &= \Cf\xf(t),
\end{aligned} \eqnlabel{filter}
\end{equation}
where, $\xf\in\Real^{\nx}$ is the state vector of the filter, $\hat{\z}\in\Real^{\nz}$ is the estimate of $\z$, and the coefficients $\Af,\Bf,\Cf$ are real matrices of appropriate dimensions to be determined.

Combining the equations \eqn{sys_proc}, \eqn{sys_meas}, \eqn{w_part} and \eqn{filter}, we get the estimation error system
\begin{equation}
\begin{aligned}
    \dot{\x}_E(t) &=  \underbrace{\begin{bmatrix} \A & \vo{0} \\ \Bf\Cy & \Af \end{bmatrix}}_{=:\Ae} \xe(t) +  \underbrace{\begin{bmatrix} \Bw \\ \Bf\Dw \end{bmatrix}}_{=:\Be} \vo{\w}(t) , \\
    \zerr(t) &= \underbrace{\begin{bmatrix} \Cz & -\Cf \end{bmatrix}}_{=:\Ce} \xe(t),
\end{aligned} \eqnlabel{est_err}
\end{equation}
\begin{align*}
    \text{where } \xe(t):=\begin{bmatrix} \x^T(t) & \xf^T(t) \end{bmatrix}^T, \text{ and }
    \zerr(t) := \z(t)-\hat{\z}(t). 
\end{align*}
The transfer matrix from $\vo{\w}(t)$ to $\zerr(t)$ for the system \eqn{est_err} is
\begin{align*}
    \GF :=  \Ce\left(s\I{}-\Ae\right)^{-1} \Be, 
\end{align*}
where $s$ is the complex variable.

Similar to \textit{Problem \ref{prob:obs}}, the filter design problem with optimal precision for a given performance $\gamma >0$ is defined next.
\begin{problem} \label{prob:fil}
  Minimize $\norm{\vo{\precs}}{1,\vo{\rho}}$ and determine feasible $\Af, \Bf, \Cf$ such that
  $\norm{\GF}{\Htwo} < \gamma$ or $\norm{\GF}{\Hinf} <\gamma$.
\end{problem}

We present the new results for $\Htwo/\Hinf$ filter design problems as theorems in the following text.

\begin{theorem}[$\Hinf$ filter] \label{thm:hinf_fil}
The solution of $\Hinf$ filter design \textit{Problem \ref{prob:fil}} is determined by solving the following optimization problem, and the filter matrices are given by $\Af = \X^{-1}\P$, $\Bf =\X^{-1}\Y$, and $\Cf = \Q$.
\begin{equation}\left.
\begin{aligned}
\min\limits_{\vo{\precs}>0,\X>0,\Y,\P,\Q,\R}\quad & \norm{\vo{\precs}}{1,\vo{\rho}} \\
\text{ such that } \M(\vo{\precs},\X,\Y,\P,\Q,\R) & < 0, \\
 \X-\R & < 0,
\end{aligned} \right\}\eqnlabel{hinf_thm_fil}
\end{equation}
where
\begin{align*}
&\M  (\vo{\precs},\X,\Y,\P,\Q,\R) := \\
& \, \begin{bmatrix} \M_{11} & \M_{12}  & \Cz^T         &  \M_{14}               & \Y\\
                 \ast     & \sym{\P} & -\Q^T         &  \M_{24}               & \Y\\
                 \ast     & \ast     & -\gamma\I{}& \vo{0}                 & \vo{0} \\
                 \ast     & \ast     & \ast          & -\gamma\I{} & \vo{0} \\
                 \ast     & \ast     & \ast          & \ast                   &-\gamma \ \diag(\vo{\precs}) \end{bmatrix}, \\
  & \M_{11}:=\sym{\R\A + \Y\Cy},
   \M_{12}:=\P + \left(\X\A + \Y\Cy\right)^T, \\
  & \M_{14}:= \R\Bd + \Y\Dd ,
  \,\, \M_{24}:= \X\Bd + \Y\Dd.
\end{align*}
\end{theorem}
\begin{IEEEproof}
We use the standard result to write the inequality $\norm{\GF}{\Hinf} <\gamma$ as the following equivalent LMIs in terms of the matrix variables $\X>0$, $\Y$, $\P$, $\Q$, and $\R$
\begin{align}
  \X-\R & < 0 \eqnlabel{temp_XR_lmi} \\
\begin{bmatrix} \M_{11}   & \M_{12}  & \R\Bw + \Y\Dw & \Cz^T  \\
                 \ast     & \sym{\P} & \X\Bw + \Y\Dw & -\Q^T  \\
                 \ast     & \ast     & -\gamma\I{}        & \vo{0} \\
                 \ast     & \ast     & \ast               & -\gamma\I{} \end{bmatrix} & < 0 , \eqnlabel{temp_lmi1}
\end{align}
where $\M_{11}:=\sym{\R\A + \Y\Cy}$ and $\M_{12}:=\P + \left(\X\A + \Y\Cy\right)^T$.
If a feasible set of matrices are found, the filter matrices are given by $\Af = \X^{-1}\P$, $\Bf =\X^{-1}\Y$, and $\Cf = \Q$ \cite{el2000advances}.

In the LMI \eqn{temp_lmi1}, using partition of matrices $\Bw$ and $\Dw$ from \eqn{w_part} yields
\begin{align*}
\begin{bmatrix} \M_{11}   & \M_{12}  & \R\Bd + \Y\Dd   & \Y\diag(\vo{\scln})   & \Cz^T  \\
                 \ast     & \sym{\P} & \X\Bd + \Y\Dd   & \Y\diag(\vo{\scln})    & -\Q^T  \\
                 \ast     & \ast     & -\gamma\I{}          & \vo{0}      & \vo{0}\\
                 \ast     & \ast     & \ast                 & -\gamma\I{} & \vo{0}\\
                 \ast     & \ast     & \ast                 & \ast        & -\gamma\I{} \end{bmatrix} < 0 .
\end{align*}
Using Schur complement,
\begin{align}
& \underbrace{\begin{bmatrix} \begin{bmatrix} \R\Bd + \Y\Dd  \\  \X\Bd + \Y\Dd \end{bmatrix} &  \begin{bmatrix} \Y \\ \Y  \end{bmatrix}\diag(\vo{\scln}) & \begin{bmatrix} \Cz^T \\ -\Q^T \end{bmatrix} \end{bmatrix}}_{=:\W} \left(\gamma\I{}\right)^{-1} \W^T  \nonumber \\
& \quad\quad\quad\quad\quad\quad\quad\quad\quad\quad\quad + \begin{bmatrix} \M_{11}   & \M_{12}  \\   \ast     & \sym{\P}  \end{bmatrix} < 0.  \eqnlabel{temp_lmi2}
\end{align}
The first quadratic term $\W\left(\gamma\I{}\right)^{-1}\W^T$  in the previous inequality can be equivalently written as $\vo{\overline{W}}\Z\vo{\overline{W}}^T$, where
\begin{align*}
 \vo{\overline{W}} &:= \begin{bmatrix}[c|c|c]
                    \Cz^T & \R\Bd + \Y\Dd & \Y \\
                    -\Q^T & \X\Bd + \Y\Dd & \Y \end{bmatrix}, \\
 \Z &:=\begin{bmatrix} \gamma^{-1}\I{} & \vo{0}                & \vo{0} \\
                        \ast                        & \gamma^{-1}\I{} & \vo{0} \\
                        \ast                        & \ast                  & \gamma^{-1}\ \diag(\vo{\scln}^2) \end{bmatrix}.
\end{align*}
Using the Schur complement to express \eqn{temp_lmi2} in terms of $\vo{\overline{W}}$ and $\Z$ gives us
\begin{align*}
  \begin{bmatrix}[c|c]
    \begin{matrix} \M_{11} & \M_{12} \\ \M_{12}^T  & \sym{\P} \end{matrix} & \vo{\overline{W}} \\ \hline
      \vo{\overline{W}}^T                                                          & - \Z^{-1}
  \end{bmatrix} < 0,
\end{align*}
or explicitly,
\begin{align}
 & \begin{bmatrix}  \M_{11} & \M_{12}  & \Cz^T       &  \M_{14}               & \Y\\
                 \ast     & \sym{\P} & -\Q^T         &  \M_{24}               & \Y\\
                 \ast     & \ast     & -\gamma\I{}& \vo{0}                 & \vo{0} \\
                 \ast     & \ast     & \ast          & -\gamma\I{} & \vo{0} \\
                 \ast     & \ast     & \ast          & \ast                   &-\gamma\ \diag(\vo{\precs}) \end{bmatrix}  <0, \eqnlabel{temp_M_lmi}
\end{align}
where $\M_{14} := \R\Bd + \Y\Dd$ and $\M_{24} := \X\Bd + \Y\Dd$.

Therefore, the solution to $\Hinf$ filter \textit{Problem \ref{prob:fil}} is determined by solving
\begin{align*}
  \min\limits_{\vo{\precs}>0,\X>0,\Y,\P,\Q,\R}\quad & \norm{\vo{\precs}}{1,\vo{\rho}}
  \text{ subject to \eqn{temp_XR_lmi} and \eqn{temp_M_lmi}}.
\end{align*}
\end{IEEEproof}

\begin{theorem}[$\Htwo$ filter] \label{thm:h2_fil}
The solution of $\Htwo$ filter design \textit{Problem \ref{prob:fil}} is determined by solving the following optimization problem, and the filter matrices are given by $\Af = \X^{-1}\P$, $\Bf =\X^{-1}\Y$, and $\Cf = \N$.
\begin{equation}\left.
\begin{aligned}
\min\limits_{\vo{\precs}>0,\X,\Y,\P,\Q,\R,\N}\quad & \norm{\vo{\precs}}{1,\vo{\rho}} \\
\text{ such that } \M(\vo{\precs},\X,\Y,\P,\R) < 0, \ \X-\R &< 0, \\
\trace{\Q} <\gamma^2, \
 \begin{bmatrix}
   -\Q & \Cz & \N \\
   \ast & -\R & -\X \\
   \ast & \ast & -\X
 \end{bmatrix} & < 0,
\end{aligned} \right\}\eqnlabel{h2_thm_fil}
\end{equation}
\begin{align*}
& \text{where } \M (\vo{\precs},\X,\Y,\P,\R) := \\
 &\qquad\qquad \begin{bmatrix} \M_{11} & \M_{12}  & \R\Bd + \Y\Dd   & \Y\\
                    \ast     & \sym{\P} & \X\Bd + \Y\Dd   & \Y\\
                    \ast     & \ast     & -\I{}            & \vo{0} \\
                    \ast     & \ast     & \ast          & -\diag(\vo{\precs})
       \end{bmatrix}, \\
  & \M_{11}:=\sym{\R\A + \Y\Cy},
   \M_{12}:=\P + \left(\X\A + \Y\Cy\right)^T.
\end{align*}
\end{theorem}
\begin{IEEEproof}
Using the standard result on the $\Htwo$ filter design, the inequality $\norm{\GF}{\Htwo} <\gamma$ is equivalently written as
\begin{align*}
 \X-\R < 0, \ \trace{\Q} <\gamma^2,
 \begin{bmatrix}
   -\Q & \Cz & \N \\
   \ast & -\R & -\X \\
   \ast & \ast & -\X
 \end{bmatrix} < 0,
\end{align*}
\begin{align}
  \begin{bmatrix} \M_{11} & \M_{12}  & \R\Bw + \Y\Dw \\
                     \ast     & \sym{\P} & \X\Bw + \Y\Dw \\
                     \ast     & \ast     & -\I{}
  \end{bmatrix} < 0 \eqnlabel{temp_M_lmi2}
\end{align}
where $\M_{11}:=\sym{\R\A + \Y\Cy}$ and $\M_{12}:=\P + \left(\X\A + \Y\Cy\right)^T$. If the LMIs are feasible, the filter matrices are given by $\Af = \X^{-1}\P$, $\Bf =\X^{-1}\Y$, and $\Cf = \N$ \cite{el2000advances}.

Similar to the previous theorem, \eqn{temp_M_lmi2} is manipulated using \eqn{w_part} and Schur complement to arrive at the optimization problem \eqn{h2_thm_fil}.
\end{IEEEproof}

We have posed the estimator design problem with optimal precision as the SDPs \eqn{hinf_thm_obs}, \eqn{h2_thm_obs}, \eqn{hinf_thm_fil} and \eqn{h2_thm_fil}.
Linear constraints on precision variables, if any, can be easily incorporated in these convex optimization problems. Standard software packages such as \texttt{CVX}\cite{grant_cvx_2020} can solve these SDPs. However, general purpose solvers do not scale well as the problem's size grows with the system dimension. We next present an ADMM algorithm to efficiently solve the precision minimization problems.

\subsection{ADMM Algorithm}

As a representative case, we present the ADMM algorithm for solving $\Hinf$ observer design problem \eqn{hinf_thm_obs}. Algorithms for other SDPs can be derived in an analogous way. A comprehensive review and tutorial of ADMM algorithms can be found in \cite{boyd_distributed_2011}.

The LMI $\M(\vo{\precs},\X,\Y)<0$ in \eqn{hinf_thm_obs} is equivalently written as
\begin{align}
  \M(\vo{\precs},\X,\Y) + \H = 0 \eqnlabel{MH_con_obs}
\end{align}
where $\H>0$ is partitioned compatibly with $\M$, i.e.
\begin{align}
  \H := \begin{bmatrix} \H_{11} & \cdots & \H_{14} \\
                        \vdots  & \ddots & \vdots \\
                      \H_{14}^T & \cdots & \H_{44} \end{bmatrix} > 0. \eqnlabel{H_part}
\end{align}
Therefore, the optimization problem \eqn{hinf_thm_obs} is re-written with the modified constraint as
\begin{equation}
\begin{aligned}
& \min\limits_{\vo{\precs}>0,\X>0,\Y,\H>0}\quad \norm{\vo{\precs}}{1,\vo{\rho}} \\ %
&\text{ such that } \M(\vo{\precs},\X,\Y) + \H = 0.
\end{aligned} \eqnlabel{hinf_obs_admm_prob}
\end{equation}

We define the \textit{augmented Lagrangian} $L_{\mu}$ for \eqn{hinf_obs_admm_prob} as
\begin{align*}
  L_{\mu}:= \norm{\vo{\precs}}{1,\vo{\rho}} + \fprod{\vo{\Lambda}}{(\M+\H)}  + (\mu/2)\fnorm{\M+\H}^2,
\end{align*}
where $\vo{\Lambda}$ is the dual variable associated with the constraint \eqn{MH_con_obs}, $\mu>0$ is the penalty parameter for the constraint violation. $\fprod{\cdot}{\cdot}$ denotes the Frobenius inner product of two matrices, and $\fnorm{\cdot}$ denotes the Frobenius norm. The augmented Lagrangian is written using the scaled dual variable $\U:=\vo{\Lambda}/\mu$ as
\begin{align}
  L_{\mu}= \norm{\vo{\precs}}{1,\vo{\rho}} + (\mu/2)\fnorm{\M(\vo{\precs},\X,\Y)+\H+\U}^2. \eqnlabel{hinf_obs_scaled_lagrange}
\end{align}

The augmented Lagrangian  $L_{\mu}$ is a function of variables $\vo{\precs},\X,\Y,\H,\U$.  The ADMM algorithm involves iterative  minimization of  $L_{\mu}$ w.r.t. each variable while holding other variables constant as shown in \eqn{hinf_obs_admm_algo}. The superscript $k+1$ in \eqn{hinf_obs_admm_algo} denotes the iteration number. For notational convenience, in \eqn{hinf_obs_admm_algo}, we show only the variable w.r.t. which minimization is to be done as the argument of $L_{\mu}(\cdot)$. It will be implied that the other variables are held constant equal to their respective latest available values. For example, in \eqn{hinf_obs_X_up}, $\X$ is the minimization variable, $\vo{\precs}$ is set to its latest value $\vo{\precs}^{k+1}$ obtained in \eqn{hinf_obs_prec_up}, and remaining variables are set to their latest values available at the end of $k^\text{th}$ iteration.
\begin{subequations}
  \begin{align}
    \vo{\precs}^{k+1} &= \arg\min \limits_{\vo{\precs}>0}  L_{\mu}(\vo{\precs}) \eqnlabel{hinf_obs_prec_up} \\
    \X^{k+1} &= \arg\min \limits_{\X>0}  L_{\mu}(\X) \eqnlabel{hinf_obs_X_up} \\
    \Y^{k+1} &= \arg\min \limits_{\Y}    L_{\mu}(\Y) \eqnlabel{hinf_obs_Y_up} \\
    \H^{k+1} &= \arg\min \limits_{\H>0}  L_{\mu}(\H) \eqnlabel{hinf_obs_H_up} \\
    \U^{k+1} &= \U^k + \M(\vo{\precs}^{k+1},\X^{k+1},\Y^{k+1}) + \H^{k+1}. \eqnlabel{hinf_obs_U_up}
  \end{align}
  \eqnlabel{hinf_obs_admm_algo}
\end{subequations}
Solutions to each optimization sub-problem in \eqn{hinf_obs_admm_algo} are presented in the appendix. 

Iterations of the ADMM algorithm are stopped when the residuals are within specified tolerances which are combinations of absolute and relative criteria \cite{boyd_distributed_2011}. Although ADMM can be slow to converge to a highly accurate solution, moderate accuracy can be achieved in reasonable number of iterations \cite{boyd_distributed_2011}. There also exist several heuristics to improve the convergence rate of the algorithm, e.g. reordering of the update steps, varying penalty parameter, etc. However, a detailed discussion on such methods is out of the scope of this paper, and an interested reader is referred to \cite{boyd_distributed_2011} and the references therein. For the purpose of numerical results discussed in the sequel, we implement the algorithm \eqn{hinf_obs_admm_algo} as presented.

\subsection{Example}
\begin{figure}[htb]
    \centering
    \includegraphics[trim={0.1cm 0.1cm 0.1cm 0.2cm},clip,width=0.47\textwidth]{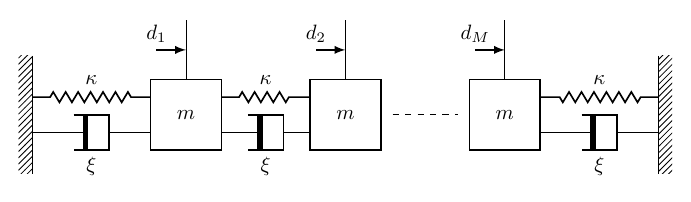} 
    \caption{Serially connected spring-mass-damper system. }
    \label{fig:SMD}
\end{figure}

We apply the proposed ADMM algorithm for the example given below. Consider an easily scalable serially connected spring-mass-damper system shown in \fig{SMD} with identical $M$ masses on a frictionless surface. Similar systems were used as test problems in previous works such as \cite{deshpande_sparseRobHinf_ACC2021, dhingra_admm_2014, zare_optimal_2018, nugroho_simultaneous_2018}.

The first and last masses in the series are attached to rigid walls. All masses $m$, spring constants $\kappa$ and damper coefficients $\xi$ are assumed to be unity. Let $x_i$ denote the distance of the \ith mass from the left wall. We define the state vector to be $\x:=[x_1,x_2,\cdots,x_M,\dot{x}_1,\dot{x}_2,\cdots,\dot{x}_M]\in\Real^{2M}$. Process noises $d_i$ act on all masses in the form of external forces. The given set of sensors measures positions and velocities of each mass. Therefore the system matrices are given by
\begin{align}
  \A =  \begin{bmatrix} \vo{0} & \I{}\\ \H & \H \end{bmatrix} ,
  \Bd =  \begin{bmatrix} \vo{0} \\ \I{}\end{bmatrix},
  \Cy = \I{}, \Dd = \vo{0},  \Cz = \I{}, \eqnlabel{ex_sys_mat}
\end{align}
where $\H$ is a tridiagonal band matrix with all principal-diagonal entries $-2$, and all super- and sub-diagonal entries $1$.

\begin{figure}[htb]
    \centering
    \includegraphics[width=0.45\textwidth]{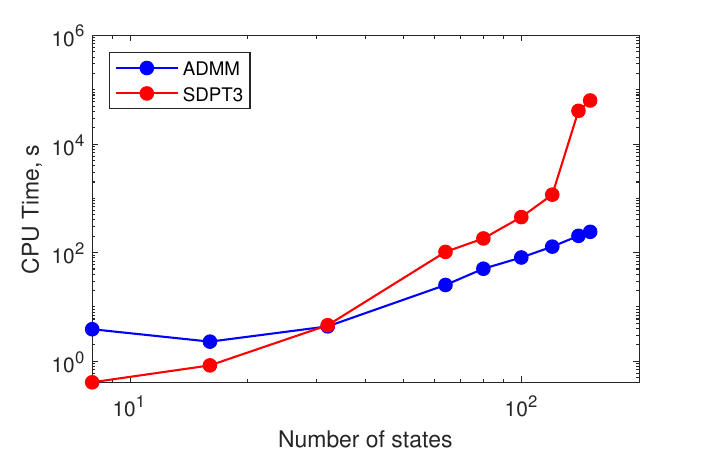}
    \caption{Variation of computation time with number of states for ADMM and  \texttt{SDPT3}. }
    \label{fig:cvx_admm_nx}
\end{figure}

The $\Hinf$ observer design problem \eqn{hinf_thm_obs} is solved using the ADMM algorithm \eqn{hinf_obs_admm_algo} with the specified performance bound $\gamma = 0.5$ and weights $\rho_i = 1$. Variation of computation (CPU) time of the ADMM algorithm with increasing number of states is shown in \fig{cvx_admm_nx}, and compared with the solver \texttt{SDPT3} \cite{toh_sdpt3_1999} which is called by the parser \texttt{CVX}\cite{grant_cvx_2020}. All numerical results presented in this paper were obtained via simulation codes implemented in MATLAB and executed on an Intel Core i5 3.4 GHz processor with 16 GB RAM.
We empirically observe that the CPU time for \texttt{SDPT3} is approximately $\mathcal{O}(\nx^6)$ for higher values of $\nx$, while the ADMM scales slightly better than $\mathcal{O}(\nx^3)$, which is consistent with the results shown in \cite{dhingra_admm_2014}. The difference between the objective values obtained using the ADMM algorithm and \texttt{CVX} is less than $1\%$ for all values of $\nx$ (not shown here).

\begin{figure}[htb]
    \centering
    \includegraphics[width=0.45\textwidth]{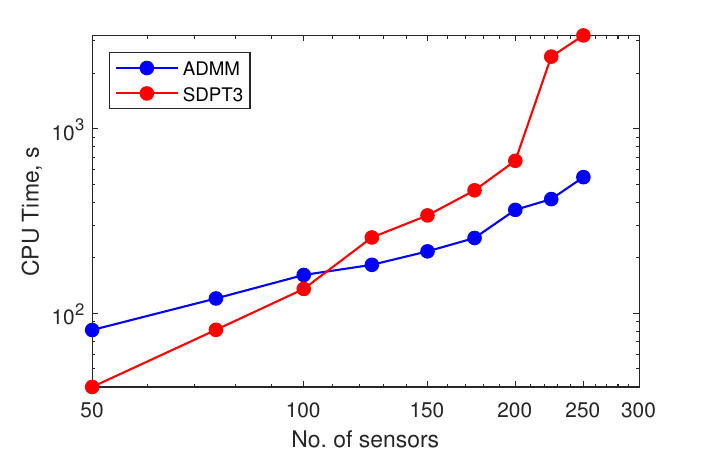}
    \caption{Variation of computation time with number of sensors for ADMM and \texttt{SDPT3}. }
    \label{fig:cvx_admm_ns}
\end{figure}

In \fig{cvx_admm_ns}, we consider another case in which we fix the number of masses $M=16$, i.e. $\nx=32$, and vary the number of sensors. The measurement matrix $\Cy$ is generated randomly while other system matrices are the same as \eqn{ex_sys_mat}. ADMM scales better than \texttt{SDPT3} as the number of sensors is increased, and ADMM's CPU time is approximately an order of magnitude smaller than \texttt{SDPT3} for large number of sensors shown in the figure.

\section{Sensor Selection Algorithms} \label{sec:sens_select}
In the previous section, we considered the first part of the problem under consideration, i.e.  $\Htwo/\Hinf$ observer and filter design to minimize precisions for a given set of sensors. In the following text, we refer to the optimization problems \eqn{hinf_thm_obs}, \eqn{h2_thm_obs}, \eqn{hinf_thm_fil} and \eqn{h2_thm_fil} as the underlying optimal precision problems or the underlying SDPs, and discuss the tractable algorithms for sensor selection.

For each of the underlying optimal precision problems, we can define a set function $f:2^{\Sens}\rightarrow \Real$, where $2^{\Sens}$ indicates the powerset of $\Sens$, such that for any $\set{Q}\subseteq\Sens$
\begin{equation} f(\set{Q}) := \left\{
  \begin{aligned}
     & \sum_{i\in\set{Q}} \rho_i\precs_i^{\ast},  \text{ if  the underlying SDP is feasible, } \\
     & +\infty, \quad \text{ otherwise,}
  \end{aligned} \right. \eqnlabel{def_f}
\end{equation}
where ${\precs_i}^{\ast}$ is the optimal solution of the SDP for the set of sensors $\set{Q}$.

Next, we note a few properties of the function $f(\cdot)$. First, by definition, $f(\cdot)$ is a nonnegative function, i.e. $f(\set{Q})\geq 0$ for each $\set{Q}\subseteq\Sens$.

It is easy to show that $f(\cdot)$ is a nonincreasing monotone, i.e. $f(\set{Q}) \geq f(\set{R})$ for any sets $\set{Q}$ and $\set{R}$ such that  $\set{Q}\subseteq\set{R}\subseteq\Sens$.
Suppose $\set{Q}\subseteq\set{R}$ and $f(\set{Q}) < \infty$. Since $\set{Q}$ is a feasible set of sensors, we can always assign arbitrarily small precisions to all sensors in $\set{R} \setminus \set{Q}$ such that $f(\set{R})$ is at max equal to $f(\set{Q})$, i.e. $f(\set{R})\leq f(\set{Q})$, which establishes the monotonicity. Another intuitive interpretation behind this property is that, the larger set $\set{R}$ will provide more degrees of freedom in the optimization problem than $\set{Q}$, and hence it will provide a better solution with smaller cost. Therefore, if there is no cardinality constraint, using all available sensors, i.e. the set $\Sens$, will yield the minimum precisions.
However, as stated in \textit{Problem \ref{prob:text}}, we are interested in identifying a smaller subset of $\Sens$ with minimal precision that satisfies the given cardinality constraint.

Using the definition \eqn{def_f}, \textit{Problem \ref{prob:text}} is written equivalently as
\begin{align}
  \min_{\set{Q}\subseteq\Sens, |\set{Q}|\leq\ks} f(\set{Q}). \eqnlabel{fmin_k}
\end{align}

Greedy algorithms have become a popular choice to solve the sensor selection problems such as \eqn{fmin_k} which involve a cardinality constraint \cite{summers_submodularity_2016, shamaiah_greedy_2010, tzoumas_sensor_2016, zhang_sensor_2017}.
The primary advantage of greedy algorithms is that they are guaranteed to yield a solution that is within the $(1-1/e)$ factor of the optimal solution in polynomial time if the objective function is submodular \cite{nemhauser_analysis_1978}. A submodular set function is defined as follows.
\begin{definition}(Submodularity) A set function $g:2^{\Sens}\rightarrow \Real$ is submodular if it satisfies
  \begin{align*}
    g(\set{R}\cup\set{Q}) + g(\set{R}\cap\set{Q}) \leq g(\set{R}) + g(\set{Q}), \ \forall \  \set{Q},\set{R}\subseteq\Sens,
  \end{align*}
and $g$ is supermodular if $-g$ is submodular.
\end{definition}

Unfortunately, in our case, the objective function in \eqn{fmin_k} is not a submodular set function, which is shown below using an example.
\begin{example}($f$ is not submodular) Consider a system with the system matrices $\Dd = \vo{0}$, $\Cz = \I{}$,
  \begin{align}
    \A = \begin{bmatrix}
     0   &  0  &   1 &   0 \\
     0  &   0 &    0 &    1 \\
    -2  &   1  &  -1  &   0 \\
     1  &  -2  &   0  &  -1
   \end{bmatrix},
   \Bd = \begin{bmatrix}
   0   &  0 \\
   0   &  0\\
   1   &  0\\
   0   &  1
 \end{bmatrix}.
  \end{align}
   The set of available sensors $\Sens=\{s_1,s_2,s_3,s_4\}$ contains element $s_i$ that measures the \ith state, i.e. the measurement matrix $\Cy=\I{}$ if all sensors are used.  We consider the optimization problem \eqn{hinf_thm_obs} with $\gamma = 0.5$, and $\rho_i = 1$.
   To show lack of submodularity, let us define $\set{Q}=\{s_1,s_4\}$ and $\set{R}=\{s_2,s_3\}$. After solving \eqn{hinf_thm_obs}, we get $f(\set{Q})=f(\set{R})=22.52$, and $f(\set{R}\cap\set{Q})=f(\emptyset)=\infty$ which clearly violates the submodularity definition. Similarly, the lack of supermodularity can be shown by defining $\set{Q}=\{s_2,s_3,s_4\}$ and $\set{R}=\{s_1,s_2,s_3\}$ that results in $f(\set{Q})=22.52$, $f(\set{R})=18.84$, $f(\set{R}\cup\set{Q})=14.0$, and  $f(\set{R}\cap\set{Q})=22.52$.
\end{example}

\begin{remark}
  Nonnegativity and nonincreasing monotonicity of $f(\cdot)$ leads to the subadditivity of $f(\cdot)$, i.e. it satisfies
   $$f(\set{R}\cup\set{Q})\leq f(\set{R}) + f(\set{Q}), \ \forall \ \set{Q},\set{R}\subseteq\Sens.$$
  Subadditivity is relatively relaxed condition than the submodularity.
\end{remark}

The subadditive property of $f(\cdot)$ is not sufficient to guarantee the performance of greedy algorithms.
The lack of submodularity of  $f(\cdot)$ implies that we may not employ classical results known for greedy algorithms to establish the optimality bounds. Nonetheless, it does not prevent us from designing and implementing greedy algorithms for solving \eqn{fmin_k}. In fact, greedy algorithms are known to perform reasonably well in practice despite the lack of submodularity \cite{zhang_sensor_2017}. Therefore, in the text to follow, we propose a novel greedy heuristic to solve \eqn{fmin_k}.

\subsection{Greedy Sensor Elimination}
In greedy sensor elimination (GSE) algorithm, instead of selecting $\ks$ sensors out of $\ns$ sensors, we eliminate $\ns-\ks$ sensors from the given set $\Sens$ so that we are effectively left with a smaller subset of $\Sens$ of cardinality $\ks$. As outlined in Algorithm \ref{algo:GSE}, we begin with the set of all sensors $\Sens$, and iteratively eliminate a sensor that results in the least increment in the cost $f(\cdot)$ until the cardinality constraint is satisfied or the algorithm reaches infeasibility. Note that in each iteration, the function $f(\cdot)$
is evaluated multiple times, i.e. the underlying optimal precision problem is solved $|\set{Q}|$ times. The total number of function evaluations of $f(\cdot)$ needed is $\frac{\ns(\ns+1)}{2}-\frac{\ks(\ks+1)}{2}$. However, multiple evaluations of $f(\cdot)$ in each iteration are amenable to parallelization which reduces the total running time of the algorithm.

\begin{algorithm}
\SetAlgoLined
 \KwInput{$\Sens$, $\vo{\rho}$, $\gamma$, $\ks$, function $f(\cdot)$ as per \eqn{def_f}}
 \KwOutput{A set $\set{Q}$ of the selected sensors}
 Initialize: $\set{Q}\leftarrow \Sens$\\
 \For{$i=1,2 \cdots \ns-\ks$}{
    \For{$s\in\set{Q}$}{
    $v_{s} := f(\set{Q}\setminus \{s\})$
  }
  $s^{\ast}:=\arg\min_{s} v_s$ \\
  \If{$v_{s^{\ast}}<\infty$}{
  $\set{Q}\leftarrow \set{Q}\setminus \{s^{\ast}\}$  \hfill {\footnotesize{// \texttt{Eliminate sensor $s^{\ast}$}}}
  }
  \Else{
  $\set{Q}\leftarrow \emptyset$; exit  \hfill {\footnotesize{// \texttt{Reached infeasibility}}}
  }
  }
 \caption{Greedy sensor elimination (GSE)}
 \label{algo:GSE}
\end{algorithm}

\subsection{Least Precise Sensor Elimination}
In \cite{li_integrating_2008}, the authors proposed a heuristic to achieve sparse sensor configuration with optimal precision. This heuristic with a slight modification to accommodate the cardinality constraint is given in Algorithm \ref{algo:LPE}.
For notational convenience, we define a set function $h:2^{\Sens}\rightarrow \Real$,
\begin{equation} h(\set{Q}) := \left\{
  \begin{aligned}
     & \{ \precs_i^{\ast}\ | \ i\in\set{Q}\},  \text{ if  the underlying SDP is feasible, } \\
     & \{ +\infty \ | \ i\in\set{Q}\}, \ \text{ otherwise,}
  \end{aligned} \right. \eqnlabel{def_h}
\end{equation}
where ${\precs_i}^{\ast}$ is the optimal solution of the SDP for the set of sensors $\set{Q}$.


\begin{algorithm}
\SetAlgoLined
 \KwInput{$\Sens$, $\vo{\rho}$, $\gamma$, $\ks$, function $h(\cdot)$ as per \eqn{def_h}}
 \KwOutput{A set $\set{Q}$ of the selected sensors}
 Initialize: $\set{Q}\leftarrow \Sens$\\
 \For{$i=1,2 \cdots \ns-\ks$}{
   $\{p_s\} = h(\set{Q})$ ;
   $s^{\ast}:=\arg\min_{s} \precs_s$ \\
  \If{$\precs_{s^{\ast}}<\infty$}{ 
  $\set{Q}\leftarrow \set{Q}\setminus \{s^{\ast}\}$  \hfill {\footnotesize{// \texttt{Eliminate sensor $s^{\ast}$}}}
  }
  \Else{
    $\set{Q}\leftarrow \emptyset$; exit  \hfill {\footnotesize{// \texttt{Reached infeasibility}}}
  }
  }
 \caption{Least precise sensor elimination (LPE)}
 \label{algo:LPE}
\end{algorithm}

The LPE algorithm can be viewed as an approximate version of the GSE algorithm wherein a sensor with the least precision is chosen for elimination instead of the one that results in the least increment in the cost. As a result, the LPE algorithm requires the underlying optimization problem to be solved $\ns-\ks$ times.

\subsection{Reweighted $l_1$-minimization}
Another widely used method for promoting sparseness is iterative reweighted $l_1$-minimization \cite{candes_enhancing_2008} which has been used for sparse sensor selection in previous works, for instance, see  \cite{jovanovic_controller_2016, munz_sensor_2014, zare_proximal_2020, deshpande_sparseH2Hinf_LCSS2021, deshpande_sparseRobHinf_ACC2021, chepuri_sparsity-promoting_2015}.
This method is adapted for the problem \eqn{fmin_k} and outlined in Algorithm \ref{algo:RLM}.

In each iteration of this algorithm, the underlying optimization problem is solved with the updated weights $\rho_s$ until sufficient number of sensors have precision within a specified tolerance $\epsilon>0$. The algorithm is said to have reached infeasibility if the maximum number of iterations $i_{\text{max}}$ is reached. Typically, the iterations converge to a sparse sensor configuration within few tens of iterations. However, that configuration may not necessarily satisfy the desired cardinality constraint.

\begin{algorithm}
\SetAlgoLined
  \KwInput{$\Sens$, $\gamma$, $\ks$, function $h(\cdot)$ as per \eqn{def_h}, $i_{\text{max}}$, $\epsilon$}
 \KwOutput{A set $\set{Q}$ of the selected sensors}
  Initialize: $i \leftarrow 1$, $\rho_s\leftarrow 1 \ \forall s\in\Sens$\\
 \While{True}{
   $\{\precs_s\} = h(\set{\Sens})$ ;
   $\set{Q}:=\{s \ | \ \precs_s>\epsilon\}$ \\
  \If{$|\set{Q}|\leq\ks$}{
    exit \hfill {\footnotesize{// \texttt{Solution found}}}
  }
  \If{$i==i_{\text{max}}$}{
  $\set{Q}\leftarrow \emptyset$; exit  \hfill {\footnotesize{// \texttt{Reached infeasibility}}}
  }
  $i\leftarrow i+1$ \\
  $\rho_s\leftarrow (\epsilon + \precs_s)^{-1} \ \forall s\in\Sens$ \hfill {\footnotesize{// \texttt{Update weights}}}
  }
 \caption{Reweighted $l_1$-minimization (RLM)}
 \label{algo:RLM}
\end{algorithm}

The performance comparison of these three algorithms is discussed next.

\subsection{Performance Comparison}
For the sake of performance comparison, we randomly generate $500$ systems using \texttt{rss} function of MATLAB \cite{mathworks_generate_nodate} with parameters $\nx = 5$, $\nd = 3$, $\ns = 12$, $\Cz = \I{}$. We consider $\Hinf$ optimal observer design problem for these systems with specified performance $\gamma=0.1$, weights $\rho = 1$ for all sensors, and impose cardinality constraint with $\ks = 4$. Therefore, to solve \eqn{fmin_k}, the algorithms solve the underlying optimization problem \eqn{hinf_thm_obs} iteratively until a solution is found or infeasibility is reached.
The reference solutions are determined via exhaustive search over all subsets of available sensors with cardinality $\ks=4$.

The performance of different algorithms is shown in Table \ref{tab:perf}.
 The first row of Table \ref{tab:perf} shows the number of random systems for which an algorithm returned a solution identical to the reference solution. The greedy algorithm (GSE) solved the optimization problem exactly for 367 out of total 500 systems, highest (by a significant margin) among the three algorithms under consideration. The second row shows the number of systems for which an algorithm reached infeasibility (incorrectly) and failed to provide a solution.  In this regard, RLM is the least reliable algorithm that could not provide a solution for 58 systems, and GSE is the most reliable algorithm that found a solution for all 500 systems.

Absolute percentage error of an algorithm for a system is defined to be $|1-\hat{f}/f^{\ast}|\times 100$, where $\hat{f}$ is the value of objective function as determined by the algorithm, and $f^{\ast}$ is the true optimal cost of the reference solution. Mean and standard deviation (SD) of absolute percentage error calculated over all systems for which an algorithm yielded a feasible solution is shown in the table. LPE is the most erroneous algorithm, while the accuracies of GSE and RLM are comparable.

Finally, the last row of Table \ref{tab:perf} shows the computational cost associated with each algorithm quantified by the number of times the underlying optimization problem \eqn{hinf_thm_obs} must be solved. For fixed $\ks$, GSE has the largest computational cost as it requires the optimization problem to be solved $\mathcal{O}(\ns^2)$ times, which is typical for greedy algorithms. On the other hand, RLM requires the  optimization problem to be solved $\mathcal{O}(10)$ times, making it the computationally cheapest algorithm. However, as noted before, GSE is the most reliable and accurate algorithm. Thus, Table \ref{tab:perf} highlights the trade-off between computational cost and reliability/accuracy of the algorithms.

\begin{table}
  \centering
  \caption{Performance comparison of different sensor selection algorithms} \label{tab:perf}
\begin{tabular}{|l|c|c|c|}
  \hline
 Algorithm & GSE (Proposed) & LPE \cite{li_integrating_2008} & RLM \cite{candes_enhancing_2008} \\
 \hline
No. of exact solutions & 367 & 203 & 276 \\
No. of infeasibilities & 0 & 1 & 58 \\
Mean \% error & 3.33\% & 270.64\% & 5.60\% \\
SD \% error & 13.03\% & 2120.73\% & 13.38\% \\
Computational cost & $\ns(\ns+1)/2$  & $\ns-\ks$ & $\mathcal{O}(10)$ \\
                   & $\quad - \ks(\ks+1)/2$ &           & \\
 \hline 
\end{tabular}
\end{table}

\section{Conclusion} \label{sec:concl}
We presented an integrated theoretical framework to design estimators (observer and filter) such that the errors are bounded by the specified $\Htwo/\Hinf$ performance criteria and the sensors precisions are minimized. We also addressed the sensor selection aspect of the problem wherein the selected set is required to satisfy a cardinality constraint. A customized ADMM algorithm was presented to solve the optimal precision problem efficiently for high-dimensional systems. We presented a new greedy algorithm for sensor selection which solves the optimal precision problem iteratively. Although the objective function for sensor selection problem was shown not to exhibit sub/super-modularity, the numerical results demonstrated that the greedy algorithm performs well in practice.
Development of a software toolbox implementing the algorithms presented in this paper for observer and filter design problems is underway.

\appendix
\subsection{ADMM algorithm for the optimal precision problem \eqn{hinf_thm_obs}} \label{app:hinf_obs}
First, note the following property of the Frobenius norm.
For any real matrix $\P$ appropriately partitioned using component matrices $\P_{ij}$, the following holds
\begin{align}
\fnorm{\P}^2 &= \sum_{i} \sum_{j} \fnorm{\P_{ij}}^2 = \norm{\vec{\P}}{2}^2, \eqnlabel{fnorm_prop}
\end{align}
where $\vec{\cdot}$ denotes the matrix vectorization operator.
Now we consider each update step in \eqn{hinf_obs_admm_algo} one by one.
\subsubsection{$\vo{\precs}$-update step}
The $\vo{\precs}$-update step in \eqn{hinf_obs_prec_up} is
 \begin{align*}
   \vo{\precs}^{k+1} &= \arg\min \limits_{\vo{\precs}>0}  \norm{\vo{\precs}}{1,\vo{\rho}} + \frac{\mu}{2}\fnorm{\M(\vo{\precs},\X^k,\Y^k)+\H^k+\U^k}^2
 \end{align*}
 Partitioning the matrix $\U^k$ compatibly similar to \eqn{H_part} and using the property \eqn{fnorm_prop}, the update becomes
\begin{align*}
  \vo{\precs}^{k+1} &= \arg\min \limits_{\vo{\precs}>0} \bigg( \norm{\vo{\precs}}{1,\vo{\rho}} + \\ & \quad\quad\quad \frac{\mu}{2}  \sum_{i} \sum_{j}  \fnorm{\M_{ij}(\vo{\precs},\X^k,\Y^k)+\H^k_{ij}+ \U^k_{ij} }^2 \bigg)\\
  &= \arg\min \limits_{\vo{\precs}>0} \left(\norm{\vo{\precs}}{1,\vo{\rho}} + \frac{\mu}{2}  \fnorm{-\gamma \ \diag(\vo{\precs})+\H^k_{44}+ \U^k_{44} }^2\right) \\
  &= \arg\min \limits_{\vo{\precs}>0} \left(\norm{\vo{\precs}}{1,\vo{\rho}} + \frac{\mu\gamma^2}{2}  \norm{\vo{\precs}-\vo{c}^k/\gamma}{2}^2\right),
\end{align*}
where $\vo{c}^k$ is the principal diagonal of  the matrix $\left(\H^k_{44}+ \U^k_{44}\right)$, and we retain the terms only which are dependent on $\vo{\precs}$. A closed form expression for $\vo{\precs}^{k+1}$ is given by
\begin{align*}
  \vo{\precs}^{k+1} = \max\left(\epsilon \, ,\,  \mathscr{S}\left(\frac{\vo{c}^k}{\gamma}, \frac{\vo{\rho}}{\mu\gamma^2}\right) \right)
\end{align*}
where the elementwise maximum operator $\max(\epsilon\, ,\, \cdot)$ projects the argument on positive orthant  approximated by a small tolerance $\epsilon>0$ such that $\vo{\precs}^{k+1}\geq\epsilon>0$, and $\mathscr{S}(\cdot \ ,\ \cdot)$ denotes the so-called \textit{soft thresholding operator} to be interpreted elementwise, and defined as
\begin{align*}
  \mathscr{S}(a , b) := \max(0 \, ,\, a-b) - \max(0 \, ,\, -a-b).
\end{align*}

\subsubsection{$\X$-update}
The $\X$-update step in \eqn{hinf_obs_X_up} is equivalent to
\begin{align*}
  \X^{k+1} = \arg\min \limits_{\X>0} \bigg( &\fnorm{\M_{11}(\X,\Y^k) + \H_{11}^k + \U_{11}^k}^2  \\ &+ 2\fnorm{\M_{12}(\X,\Y^k) + \H_{12}^k + \U_{12}^k}^2 \bigg),
\end{align*}
where we have used the property \eqn{fnorm_prop} again, and retained the terms only which depend on $\X$.
Using the definitions of $\M_{11}$ and $\M_{12}$  from \eqn{hinf_obs_m11m12_def}
\begin{align*}
  \X^{k+1} = \arg\min \limits_{\X>0} \bigg(&\fnorm {\X\A + \A^T\X + \vo{V}_{11}^k}^2 \\& \quad\quad  + 2\fnorm{\X\Bd + \vo{V}_{12}^k}^2\bigg),
\end{align*}
where
\begin{align*}
  \vo{V}_{11}^k & := \sym{\Y^k\Cy}+\H_{11}^k + \U_{11}^k ,\\
  \vo{V}_{12}^k & := \Y^k\Dd+\H_{12}^k + \U_{12}^k.
\end{align*}
Let us denote $\vo{v}_{11}^k:= \vec{\vo{V}_{11}^k}$ and $\vo{v}_{12}^k:= \vec{\vo{V}_{12}^k}$. Then using  \eqn{fnorm_prop} and the identity $\vec{\A\B\C} = (\C^T\otimes\A)\vec{\B}$, we get
\begin{align}
  \X^{k+1} &= \arg\min \limits_{\X>0} \bigg( \norm{ (\A^T\otimes\I{\nx} +  \I{\nx}\otimes\A^T) \vec{\X}  +       \vo{v}_{11}^k}{2}^2 \nonumber \\
    & \quad\quad\quad\quad  + 2\norm{(\Bd^T\otimes\I{\nx})\vec{\X} + \vo{v}_{12}^k}{2}^2\bigg), \nonumber\\
   &= \arg\min \limits_{\X>0} \left( \norm{\vo{\overline{A}} \,\vec{\X} +  \vo{\overline{v}}^k}{2}^2 \right), \eqnlabel{hinf_X_up_prob}
\end{align}
where,
\begin{align*}
  \vo{\overline{A}}:= \begin{bmatrix} (\A^T\otimes\I{\nx} +  \I{\nx}\otimes\A^T) \\ \sqrt{2}(\Bd^T\otimes\I{\nx}) \end{bmatrix}, \vo{\overline{v}}^k := \begin{bmatrix} \vo{v}_{11}^k \\ \sqrt{2} \vo{v}_{12}^k \end{bmatrix},
\end{align*}
and for unambiguity, in this section we denote the identity matrix of dimension $N$ by $\I{N}$.

Equation \eqn{hinf_X_up_prob} is the least squares problem subject to the constraint $\X>0$. We again implement an \textit{inner loop} of the ADMM algorithm to solve \eqn{hinf_X_up_prob} as discussed in Appendix \ref{sec:admm_inner} and obtain $\X^{k+1}$.

One benefit of using ADMM is that the inner loops such as \eqn{hinf_admm_innerX} can be terminated prematurely, i.e. the outer ADMM loop \eqn{hinf_obs_admm_algo} converges to a solution with moderate accuracy even if the optimization sub-problems such as \eqn{hinf_obs_X_up} are not solved exactly \cite{candes_enhancing_2008}. Therefore, it is also possible to approximate the solution of \eqn{hinf_X_up_prob} without implementing the inner loop \eqn{hinf_admm_innerX}.

The iterate $\X^{k+1}$ is approximated by solving the least squares problem by constraining $\X$ to be symmetric, i.e. $\X=\X^T$ but not positive definite, and then the solution of the relaxed least squares problem is projected on the positive definite cone, i.e.
\begin{subequations}
  \begin{align}
    \X^{k+1} & \approx \mathscr{P} \left(\arg\min \limits_{\X=\X^T} \left( \norm{\vo{\overline{A}} \,\vec{\X} +  \vo{\overline{v}}^k}{2}^2 \right) \right)  \\
    & = \mathscr{P}  \left(\arg\min \limits_{\X=\X^T} \left( \norm{\vo{\overline{A}}_r \,\textbf{vec}_r(\X) +  \vo{\overline{v}}^k}{2}^2 \right) \right) \eqnlabel{hinf_X_up_prob_red1} \\
    &=  \mathscr{P} \left( \X_{ls} \right) \text{ such that } \textbf{vec}_r(\X_{ls}) = -\vo{\overline{A}}_r^{\dagger}\vo{\overline{v}}^k,  \eqnlabel{hinf_X_up_prob_red2}
  \end{align}\eqnlabel{hinf_X_up_prob_red}
\end{subequations}

where $\mathscr{P}(\cdot)$ is a projection operator which projects the argument on the cone of  positive definite matrices. For a given symmetric matrix $\P$, let its eigenvalue decomposition is given by $\P = \R \ \diag(\vo{\lambda})\R^T$ where $\vo{\lambda}$ are eigenvalues and $\R$ is the matrix of eigenvectors. Then the projection of $\P$ on positive definite cone is given by
\begin{align}
  \mathscr{P}(\P):= \R \ \diag\left(\max(\epsilon \, ,\, \vo{\lambda})\right) \R^T \eqnlabel{sdp_project}
\end{align}
where the  positive definite cone is approximated by a small tolerance $\epsilon>0$ such that $\P\geq\epsilon\I{}>0$.

The constraint $\X=\X^T$ reduces the dimension of the least squares problem in \eqn{hinf_X_up_prob_red},
and hence written in terms of $\textbf{vec}_r(\X)$ which is the vector of unique entries of $\vo{{X}}$ (i.e. lower triangular elements), and the \text{reduced} matrix $\vo{\overline{A}}_r$   is obtained by combining appropriate columns of $\vo{\overline{A}}$. The least squares solution $\X_{ls}$ is obtained using pseudo-inverse of $\vo{\overline{A}}_r$.

 The update equation \eqn{hinf_X_up_prob_red} can be viewed as a single iteration of the inner ADMM loop \eqn{hinf_admm_innerX}.

\subsubsection{$\Y$-update}
Similar to the $\X$-update, the $\Y$-update step in \eqn{hinf_obs_Y_up} is written as
\begin{align*}
  \Y^{k+1} =  \arg\min \limits_{\Y} \bigg( &\fnorm{\M_{11}(\X^{k+1},\Y) + \H_{11}^k + \U_{11}^k}^2  \\ &+ 2\fnorm{\M_{12}(\X^{k+1},\Y) + \H_{12}^k + \U_{12}^k}^2 \\  &  + 2\fnorm{\Y+\H_{14}^k + \U_{14}^k}^2\bigg),
\end{align*}
\begin{align}
  \Y^{k+1} &=  \arg\min \limits_{\Y} \left( \norm{\vo{\overline{C}} \,\vec{\Y} + \vo{\overline{z}}^k}{2}^2 \right)  ,\eqnlabel{hinf_Y_up_prob}
\end{align}
where
\begin{align*}
  \vo{\overline{C}}&:= \begin{bmatrix} \left(\Cy^T\otimes\I{\nx} +  (\I{\nx}\otimes\Cy^T)\mathcal{T} \right)\\ \sqrt{2} \ (\Dd^T\otimes\I{\nx})  \\ \sqrt{2} \ \I{(\nx\ny)} \end{bmatrix},
\end{align*}
\begin{align*}
  \vo{\overline{z}}^k &:= \begin{bmatrix} \vec{\X^{k+1}\A + \A^T\X^{k+1} + \H_{11}^k + \U_{11}^k} \\ \sqrt{2} \ \vec{\X^{k+1}\Bd + \H_{12}^k + \U_{12}^k} \\  \sqrt{2} \ \vec{ \H_{14}^k + \U_{14}^k} \end{bmatrix} ,
\end{align*}
and $\mathcal{T} \in \Real^{\nx\ny \times \nx\ny}$ denotes a linear transformation operator matrix such that $\vec{\Y^T} = \mathcal{T} \ \vec{\Y}$.
Solution  to the least squares problem \eqn{hinf_Y_up_prob} is simply $\vec{\Y^{k+1}} = - \vo{\overline{C}}^{\dagger} \vo{\overline{z}}^k$.

\subsubsection{$\H$-update}
The $\H$-update step \eqn{hinf_obs_H_up} takes the following simple form
\begin{align*}
  \H^{k+1} &= \arg\min \limits_{\H>0}   \fnorm{\M(\betab^{k+1},\X^{k+1},\Y^{k+1})+\H+\U^k}^2 \\
           &= \mathscr{P}\left( -\M(\betab^{k+1},\X^{k+1},\Y^{k+1}) - \U^k \right),
\end{align*}
where the projection operator $ \mathscr{P}(\cdot)$ is defined in \eqn{sdp_project}.

\subsection{Solution of \eqn{hinf_X_up_prob} using ADMM} \label{sec:admm_inner}
We re-write \eqn{hinf_X_up_prob} as
\begin{align}
  \min \limits_{\vo{\hat{X}}=\vo{\hat{X}}^T,\vo{\hat{Z}}>0} \left( \norm{\vo{\overline{A}} \,\vec{\vo{\hat{X}}} +  \vo{\overline{v}}^k}{2}^2 \right) \text{ s.t. } \vo{\hat{X}}-\vo{\hat{Z}}=0.
  \eqnlabel{hinf_X_up_prob_admm}
\end{align}
wherein we denote the variables involved in the inner loop with an overhead hat to differentiate them from the outer loop variables in \eqn{hinf_obs_admm_algo}.

 The augmented Lagrangian for \eqn{hinf_X_up_prob_admm} in terms of the scaled dual variable $\vo{\hat{U}}$ is
 \begin{align*}
   \hat{L}_{\hat{\mu}}:=&  \norm{\vo{\overline{A}} \,\vec{\vo{\hat{X}}}+\vo{\overline{v}}^k}{2}^2  + (\hat{\mu}/2)\fnorm{\vo{\hat{X}}-\vo{\hat{Z}}+\vo{\hat{U}}}^2 \\
    =&\norm{\vo{\overline{A}} \,\vec{\vo{\hat{X}}}+\vo{\overline{v}}^k}{2}^2  + (\hat{\mu}/2)\norm{\vec{\vo{\hat{X}}- \vo{\hat{Z}}+\vo{\hat{U}}}}{2}^2
 \end{align*}
Therefore, the ADMM algorithm involves the following steps
\begin{subequations}
  \begin{align}
    \vo{\hat{X}}^{j+1} &= \arg\min \limits_{\vo{\hat{X}}=\vo{\hat{X}}^T}   \hat{L}_{\hat{\mu}}(\vo{\hat{X}},\vo{\hat{Z}}^j,\vo{\hat{U}}^j) \eqnlabel{hinf_X_up_innerX} \\
    \vo{\hat{Z}}^{j+1} &= \arg\min \limits_{\vo{\hat{Z}}>0} \hat{L}_{\hat{\mu}}(\vo{\hat{X}}^{j+1},\vo{\hat{Z}},\vo{\hat{U}}^j)  \nonumber \\
                      &=  \mathscr{P}\left( \vo{\hat{X}}^{j+1} + \vo{\hat{U}}^j \right) \eqnlabel{hinf_Z_up_innerX} \\
    \vo{\hat{U}}^{j+1} &= \vo{\hat{U}}^j + \vo{\hat{X}}^{j+1} - \vo{\hat{Z}}^{j+1} \eqnlabel{hinf_U_up_innerX}
  \end{align}
  \eqnlabel{hinf_admm_innerX}
\end{subequations}
where the projection operator $ \mathscr{P}(\cdot)$ is defined in \eqn{sdp_project}.
The step \eqn{hinf_X_up_innerX} is equivalent to
\begin{align*}
  \vo{\hat{X}}^{j+1} &= \arg\min \limits_{\vo{\hat{X}}=\vo{\hat{X}}^T}   \hat{L}_{\hat{\mu}}(\vo{\hat{X}},\vo{\hat{Z}}^j,\vo{\hat{U}}^j) \\
      &= \arg\min \limits_{\vo{\hat{X}}=\vo{\hat{X}}^T} \left( \norm{\vo{\hat{A}} \,\vec{\vo{\hat{X}}} +  \vo{\hat{v}}^{k,j}}{2}^2 \right)  \\
      &= \arg\min \limits_{\vo{\hat{X}}=\vo{\hat{X}}^T}\left( \norm{\vo{\hat{A}}_r \,\textbf{vec}_r(  \vo{\hat{X}}) +  \vo{\hat{v}}^{k,j}}{2}^2 \right) \\
      & = \vo{\hat{X}}_{ls} \text{ such that } \textbf{vec}_r(\vo{\hat{X}}_{ls}) = -\vo{\hat{A}}_r^{\dagger}\vo{\hat{v}}^{k,j},
\end{align*}
where
\begin{align*}
  \vo{\hat{A}}:= \begin{bmatrix} \vo{\overline{A}} \\ \sqrt{\hat{\mu}/2} \ \I{\nx^2} \end{bmatrix}, \vo{\hat{v}}^{k,j} := \begin{bmatrix} \vo{\overline{v}}^k \\ \sqrt{\hat{\mu}/2} \ \vec{ -\vo{\hat{Z}}^j + \vo{\hat{U}}^j } \end{bmatrix},
\end{align*}
 and the least squares solution is obtained similar to \eqn{hinf_X_up_prob_red1} and \eqn{hinf_X_up_prob_red2}.


\bibliographystyle{unsrt}
\bibliography{MyLibrary}

%

\end{document}